\documentclass[aps,prl,reprint,groupedaddress]{revtex4-2}

\usepackage{graphicx}
\usepackage{amsmath}
\usepackage{amssymb}

\begin{document}

\title{Coupled surface and bulk diffusion in crystals}


\author{S.\,S. Kosolobov}
\email[]{s.kosolobov@skoltech.ru}
\affiliation{Skolkovo Institute of Science and Technology, Bolshoy Boulevard 30, bld. 1, Moscow 121205, Russia}


\date{\today}

\begin{abstract}
We analyze point defect bulk and surface diffusion near the crystal-vacuum interface and show that bulk diffusion is altered by atomic reactions at the surface-vacuum boundary. A new atomic mechanism for point defect generation and recombination is presented. In this mechanism, atomic steps are considered as sources and sinks not only for adsorbed atoms and surface vacancies but also for bulk native point defects --- vacancies and self-interstitials. We demonstrate that bulk diffusion is coupled with the diffusion of adatoms and surface vacancies at the surface via the atomic processes in the intermediate subsurface layer, containing adsorbed self-interstitials and bulk vacancies. The results show the existence of the fundamental relation between equilibrium concentrations of the point defects in bulk and at the surface of the crystal.   
\end{abstract}


\maketitle


In crystalline systems, point defects such as interstitials and vacancies play a significant role in controlling physical properties of the material and device performance~\cite{pantelides1978electronic,yoshida2015defects,koenraad2011single}. Intrinsic point defects critically affect the electronic, optical, and magnetic properties of semiconductors~\cite{park2018point,watanabe2004direct}, they play a significant role in heterogeneous catalysis~\cite{boyes1985cathodoluminescence,ledentu2000heterogeneous}, ion-implantation and radiation-induced processes~\cite{toijer2021solute,zhang2018radiation}, and affect physical properties of semiconductor nanostructures~\cite{holmberg2013imaging,sakurai2012role}. As the technology scales down to nanoscale, the role of surfaces and interfaces becomes increasingly important. In general, surfaces usually are considered as perfect sinks and sources for point defects~\cite{beyerlein2015defect}. However, despite the importance of defect interactions with surfaces and interfaces, the understanding of the atomic mechanisms of these interactions is still questionable. Here we show that bulk diffusion is coupled with the diffusion of adatoms and surface vacancies at the surface via the atomic processes in the intermediate subsurface layer, containing adsorbed self-interstitials and bulk vacancies. The results show the existence of the fundamental relation between equilibrium concentrations of the point defects in bulk and at the surface of the crystal. Our results demonstrate how the crystal surface maintain the equilibrium concentrations of point defects at the crystal boundary by producing or adsorbing the bulk vacancies and self-interstitial by atomic steps. Our study provides evidence  for the link between surface and bulk diffusion processes and increases the understanding  of the point defect behavior in crystals and nanoscale structures in various solid-state systems.   

Bulk self- and dopant diffusion studies provide valuable information about model parameters entering diffusion equations, describing the diffusion and interaction of point defects. The solution of differential equations depends on the boundary conditions. In particular, in widely used Dirichlet and Neumann type of boundary conditions the equilibrium concentrations of point defects are the critical parameters, that affect the solution of the diffusion equation~\cite{pichler_book2004}. However, it is not clear how surface maintains the equilibrium concentration of bulk point defects. Thus, it is of critical importance to understand the nature of the diffusion and atomic mechanisms of the point defect generation and recombination in the vicinity of crystal boundaries, which control the properties of solids.

\begin{figure} [b]
	\includegraphics[width=7.5cm]{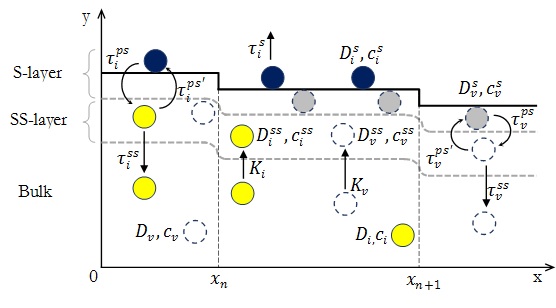}
	\caption{\label{fig-bulk-model} Schematic representation (color online) of the model with surface layer (S-layer), subsurface layer (SS-layer) and bulk displaying mechanism of point defect formation, diffusion and annihilation near the crystal-vacuum boundary. Dark and gray circles represent adatoms and surface vacancies, respectively, yellow and white circles are self-interstitials and bulk vacancies. }
\end{figure}

In this Letter, we focus on native point defect behaviour near the surface boundary. We extend the model proposed in Ref.~\cite{kosolobov2019subsurface} to include the bulk point defect diffusion and interaction with the crystal boundary. Figure~\ref{fig-bulk-model} represents the proposed model of the crystal.  The surface layer denoted as $S$-layer contains singular terraces, divided by straight atomic steps in the positions $x_{n}$ and $x_{n+1}$. Adsorbed atoms and surface vacancies at the crystal surface are represented by dark solid and gray dashed circles, respectively. 
First, we consider the process of the bulk native point defect generation, diffusion, and recombination.
To simplify the problem we ignore any internal sources and sinks for point defects for the moment.

We use the system of reaction-diffusion equations (\ref{eq:system-bulk-i}),(\ref{eq:system-bulk-v}),
that govern bulk point defect concentrations~\cite{fahey1989,hu1994nonequilibrium}. The first term in the Eqs.~(\ref{eq:system-bulk-i}), and (\ref{eq:system-bulk-v}) describe the bulk diffusion of the self-interstitials and vacancies, denoted by subscripts $i$ and $v$, respectively. The diffusion coefficients $D_{i}$, $D_{v}$ assumed not to be dependent on the concentrations of self-interstitials $c_{i}$ and vacancies $c_{v}$. 

\begin{eqnarray}
	\label{eq:system-bulk-i}
	\frac{\partial c_{i}}{\partial t} &= D_{i}\frac{\partial^2 c_{i}}{\partial y^2} - k_{bm}\left(c_{i}c_{v}-c_{i}^*c_{v}^*\right), \\
	\label{eq:system-bulk-v}
	\frac{\partial c_{v}}{\partial t} &= D_{v}\frac{\partial^2 c_{v}}{\partial y^2} - k_{bm}\left(c_{i}c_{v}-c_{i}^*c_{v}^*\right). 
\end{eqnarray}

The second term describes the net rate of Frenkel pair generation-recombination, where $c_{i}^{*}$ and $c_{v}^{*}$ denote equilibrium concentrations of the corresponding point defects. The reaction rate constant can be expressed as $k_{bm}=4 \pi r_{i} (D_{i}+D_{v})$, where $r_{i}$ has the order of the interatomic distance in the crystal~\cite{mikelsen2005,watkins2008vacancy,bracht2003radiation,kissinger2014simulation}.

Atomic reactions of the self-interstitials and bulk vacancies at the crystal-vacuum boundary are described by kinetic coefficients $K_{i,v}=D_{i,v}/\Lambda_{i,v}=a\nu \cdot exp\left( -E_{i,v}^{ads}/kT \right)$, where $\Lambda_{i,v}$ have the same meaning as the adatom kinetic length at the crystal surface, introduced for the description of the adatom-step interaction in~\cite{man2007kinetic,bales1997}, $a$ denotes the elemental diffusion step (the distance between two neighboring equilibrium positions for the point defect) in the bulk, $\nu$ --- atomic frequency, and $E_{i,v}^{ads}$ is the energy barrier for the adsorption of the point defect from the bulk to intermediate subsurface layer ($SS$-layer).

The subsurface layer contains the self-interstitials (yellow circles) and bulk vacancies (dashed white circles) adsorbed just below the crystal-vacuum interface (Fig.~\ref{fig-bulk-model})~\cite{kosolobov2019subsurface}. Adsorbed bulk point defects diffuse along the surface boundary in the $SS$-layer with the diffusion coefficients $D_{i}^{ss}$ and $D_{v}^{ss}$, respectively. The concentrations of the self-interstitials $c_{i}^{ss}$ and vacancies $c_{v}^{ss}$ are defined by the processes of the generation and recombination of the defects at the atomic steps. These are quite natural processes because adding or removing the defect at the edge of the atomic step corresponds to the motion of the step, leaving the surface energy unchanged. We propose that adatoms and advacancies can penetrate to $SS$-layer and emerge to the surface with the probabilities, defined by the corresponding lifetimes $\tau_{i,v}^{ps}$ and $\tau_{i,v}^{ps\prime}$.

Next we consider the surface point defects diffusing on a terrace. The surface concentration $c_{i}^{s}$ for adatoms and $c_{v}^{s}$ for advacancies are governed by the system of the continuity equations~(\ref{eq:c-i-s}), (\ref{eq:c-v-s}). The first term on the right hand side in both equations corresponds to the surface diffusion of the point defect on the trace of width $L$, where $D_{i}^{s}$ and $D_{v}^{s}$ are diffusion coefficients for adatoms and advacancies, respectively.  

\begin{eqnarray}
	\label{eq:c-i-s}
	\frac{\partial c^{s}_{i}}{\partial t} = D_{i}^{s}\frac{\partial^2 c^{s}_{i}}{\partial x^2} - k^{s}_{bm}\left(c_{i}^{s}c_{v}^{s}-c_{i}^{s^*}c_{v}^{s^*}\right) \nonumber \\
	-\frac{c_{i}^{s}}{\tau_{i}^{s}} -\frac{c_{i}^{s}}{\tau_{i}^{ps}}+\frac{c_{i}^{ss}}{\tau_{i}^{ps\prime}} , \\
	\label{eq:c-v-s}
	\frac{\partial c^{s}_{v}}{\partial t} = D_{v}^{s}\frac{\partial^2 c^{s}_{v}}{\partial x^2} - k^{s}_{bm}\left(c_{i}^{s}c_{v}^{s}-c_{i}^{s^*}c_{v}^{s^*}\right) \nonumber \\
	-\frac{c_{v}^{s}}{\tau_{v}^{ps}}+\frac{c_{v}^{ss}}{\tau_{v}^{ps\prime}}. 
\end{eqnarray}

The second term in both equations are referred to the generation-recombination of the adatom-advacancy pairs on the terrace, where $c_{i,v}^{s^*}$ denote the equilibrium concentrations of adatoms and advacancies, and $k^{s}_{bm}$ describes the net rate of the generation-recombination of the surface point defects.  Adatom desorption from the surface due to the sublimation process is described by the third term in Eq.~(\ref{eq:c-i-s}), where $\tau_{i}^{s}$ is the adatom lifetime. The last two terms in Eqs.~(\ref{eq:c-i-s}), (\ref{eq:c-v-s}) describe the point defect exchange between surface and subsurface layers. The corresponding lifetimes $\tau_{i,v}^{ps}$ are referred to the penetration of the surface point defects into the subsurface layer, whereas the reverse process of floating the interstitials and vacancies from the subsurface to the surface is described by $\tau_{i,v}^{ps\prime}$.
The boundary conditions which reflect the finite rates of the atomic reactions of the surface point defects at the step edges are determined by

\begin{equation}
	-D_{i,v}^{s}\left.\frac{\partial c_{i,v}^{s}}{\partial x}\right|_{x_{n},x_{n+1}} = k_{i,v}^{s}\left(c_{i,v}^{s^*}-\left.c_{i,v}^{s}\right|_{x_{n},x_{n+1}}\right),
\end{equation}
  
\noindent where $c_{i,v}^{s^*}, c_{i,v}^{s^*}$ are the equilibrium concentrations of adatoms and advacancies, $k_{i,v}^{s}$ are the kinetic coefficients that define the interaction rate of the adsorbed species with atomic step and are introduced in analogy with $\beta_{s}$ from~\cite{bales1997}.

The continuity equations, describing the bulk point defect diffusion along the crystal-vacuum boundary are given as follows:

	\begin{eqnarray}
	\label{eq:c-i-ss}
		\frac{\partial c_{i}^{ss}}{\partial t} = D_{i}^{ss}\frac{\partial^2 c_{i}^{ss}}{\partial x^2}-k^{ss}_{bm}\left(c_{i}^{ss}c_{v}^{ss}-c_{i}^{ss^*}c_{v}^{ss^*}\right) \nonumber \\
		+K_{i}\left.c_{i} \right|_{y=H} -\frac{c_{i}^{ss}}{\tau_{i}^{ss}}+\frac{c_{i}^{s}}{\tau_{i}^{ps}}-\frac{c_{i}^{ss}}{\tau_{i}^{ps\prime}}, \\
	\label{eq:c-v-ss}
		\frac{\partial c_{v}^{ss}}{\partial t} = D_{v}^{ss}\frac{\partial^2 c_{v}^{ss}}{\partial x^2}-k^{ss}_{bm}\left(c_{i}^{ss}c_{v}^{ss}-c_{i}^{ss^*}c_{v}^{ss^*}\right) \nonumber \\
		+K_{v}\left.c_{v} \right|_{y=H} -\frac{c_{v}^{ss}}{\tau_{v}^{ss}}+\frac{c_{v}^{s}}{\tau_{v}^{ps}}-\frac{c_{v}^{ss}}{\tau_{v}^{ps\prime}}.  
	\end{eqnarray} 
 
\noindent Here the first and the second term on the right hand side of equations~(\ref{eq:c-i-ss}), and (\ref{eq:c-v-ss}) refer to lateral diffusion of the self-interstitials and vacancies in the subsurface layer along the crystal-vacuum boundary and generation-recombination of the point defects in the subsurface layer, respectively. The third term in both equations~(\ref{eq:c-i-ss}) and (\ref{eq:c-v-ss}) represents the flux of the corresponding bulk point defects from the bulk to the subsurface adsorbed layer. The fourth term describes the variation of the defect concentrations in the SS-layer due to "desorption" of the corresponding point defect to the bulk.  So defect concentration in the SS-layer depends on the supersaturation of the self-interstitials and vacancies in the bulk, that defined by Eqs.~(\ref{eq:system-bulk-i}), and (\ref{eq:system-bulk-v}). And the last two cross-terms are linking equations (\ref{eq:c-i-s}),\,(\ref{eq:c-v-s}) and (\ref{eq:c-i-ss}),\,(\ref{eq:c-v-ss}). They describe the exchange processes of defects among surface and subsurface layers.   

The detachment of the point defects from the $SS$-layer to the bulk is characterized by the lifeitimes $\tau_{i}^{ss}=\nu^{-1}\cdot exp \left(E_{i}^{p}/kT\right)$, $\tau_{v}^{ss}=\nu^{-1}\cdot exp \left(E_{v}^{p}/kT\right)$ for self-interstitials and vacancies, respectively. Here $E_{i,v}^{p}$ are energy barriers for detaching the native point defects from the subsurface layer to the bulk.     
Eqations (\ref{eq:c-i-ss}), and (\ref{eq:c-v-ss}) are supplemented by boundary conditions:

	\begin{equation}
		-D_{i,v}^{ss}\left.\frac{\partial c_{i,v}^{ss}}{\partial x}\right|_{x_{n},x_{n+1}} = k_{i,v}^{ss}\left(c_{i,v}^{ss^*}-\left.c_{i,v}^{ss}\right|_{x_{n},x_{n+1}}\right),
	\end{equation}
	
\noindent where $D_{i}^{ss}$ and $D_{v}^{ss}$ are diffusion coefficients, describing lateral diffusion in the SS-layer for self-interstitials and vacancies, respectively. We introduce the equilibrium concentrations of the self-interstitials $c_{i}^{ss^*}$, and vacancies $c_{v}^{ss^*}$ in the subsurface layer and the kinetic coefficients $k_{i,v}^{s}$ which describe the interaction of the diffusing defects with the atomic step. 

Finally, we obtain the system of self-consistent differential equations~(\ref{eq:system-bulk-i})-(\ref{eq:c-v-s}), (\ref{eq:c-i-ss})-(\ref{eq:c-v-ss}), describing the generation, recombination, and diffusion of the point defects in the bulk, subsurface layer, and at the surface. The initial conditions are taken for the equilibrium case: $c_{i,v}=c_{i,v}^{*}$, $c_{i,v}^{s}=c_{i}^{s*}$, and $c_{i,v}^{ss}=c_{i}^{ss*}$.
In total equilibrium, one can expect the equilibration of all fluxes between surface and subsurface layers: $	c_{i,v}^{s^*}/\tau_{i,v}^{ps}=c_{i,v}^{ss^*}/\tau_{i,v}^{ps\prime}$ and between subsurface layer and bulk: $c_{i,v}^{ss^*}/\tau_{i,v}^{ss}=K_{i,v} c_{i,v}^{*}$. Thus, we can find the relations between the equilibrium concentrations of the point defects: $c_{i,v}^{s^*}=\left(\tau_{i,v}^{ps}/\tau_{i,v}^{ps\prime}\right)c_{i,v}^{ss^*}$ and $c_{i,v}^{ss^*}=K_{i,v}  \tau_{i,v}^{ss} c_{i,v}^{*}$.
Substituting the kinetic coefficients $K_{i,v}$ and the corresponding lifetimes $\tau_{i,v}^{ps}=\nu^{-1} exp\left( E_{i,v}^{ps}/kT \right)$, $\tau_{i,v}^{ps\prime}=\nu^{-1} exp\left( E_{i,v}^{ps \prime}/kT \right)$ in the last relations leads to:

\begin{equation}
	\label{eq:Ei-ads}
	E_{i,v}^{ads}-E_{i,v}^{p} = E_{i,v}^{ps}-E_{i,v}^{ps \prime} + kT \cdot Ln\left(a \frac{c_{i,v}^*}{c_{i,v}^{s^*}} \right) \\
\end{equation}

\noindent Equation~(\ref{eq:Ei-ads}) reflects the relation between the equilibrium concentrations of the point defects (adatoms $c_{i}^{s^*}$, advacancies $c_{v}^{s^*}$, bulk self-interstitials $c_{i}^{*}$, bulk vacancies $c_{v}^{*}$), and corresponding energy barriers. Note, if the energy barriers $E_{i,v}^{ads} \neq E_{i,v}^{p}$ and $E_{i,v}^{ps} \neq E_{i,v}^{ps \prime}$, than we have $c_{i,v}^* \neq c_{i,v}^{s^*}/a$.   

In general, the equilibrium concentrations of the point defects are given by $c_{i,v}^*=n_s \times exp\left( -E_{i,v}^{f}/kT \right)$, $c_{i,v}^{s^*}=n_{ss} \times exp\left( -E_{i,v}^{ad}/kT \right) $, where $n_s, n_{ss}$ denote the bulk and surface atom density, $E_{i,v}^{f}$ are the formation energy of the corresponding defect and $E_{i,v}^{ad}$ refer to the activation energy for the surface defect formation.   
Additionally, in simple cubic Kossel-Stranski model $n_{ss}/a=n_s$. Thus, equation~(\ref{eq:Ei-ads}) transforms to:

\begin{equation}
	\label{Energylaw}
	 (E_{i,v}^{ps}+E_{i,v}^{p})-(E_{i,v}^{ads}+E_{i,v}^{ps \prime}) = (E_{i,v}^{f} - E_{i,v}^{ad}),   \\
\end{equation}

\noindent Here the sum of the energy barriers $E_{i,v}^{p} + E_{i,v}^{ps}=E_{i,v}^{p+ps}$ restricts the flux of the point defect from surface to bulk preventing the dissolving of point defects. The second term on the left hand side of the equation $E_{i,v}^{ads} + E_{i,v}^{ps \prime}=E_{i,v}^{ads+ps\prime}$ represents the energy barrier that limits the reverse fluxes of the point defects from bulk to the surface.  
We find that the difference between energy barriers for penetration and emerging of the point defects $E_{i,v}^{p+ps} - E_{i,v}^{ads+ps\prime}$ through the surface boundary is constant and defined by the formation energies $E_{i,v}^{f}$ and $E_{i,v}^{ad}$ for the bulk and surface point defects, respectively. It is noted that obtained relation (Eq.~(\ref{Energylaw})) does not depend on temperature.

Equation.~(\ref{Energylaw}) describes a connection between the processes of point defect formation in bulk and at the surface of the crystal. It is seen that if there is no energy barrier between surface and bulk ($E_{i,v}^{p}-E_{i,v}^{ads}=0$ and $E_{i,v}^{ps}-E_{i,v}^{ps \prime}=0$),  then, according to Eq.(\ref{eq:Ei-ads}), $c_{i,v}^*=c_{i,v}^{s^*}/a$. Or, in other words, by equalizing the corresponding energy barriers $E_{i,v}^{ads}=E_{i,v}^{p}$ and $E_{i,v}^{ps}=E_{i,v}^{ps \prime}$ we equalize the equilibrium concertations of the point defects at the surface $c_{i,v}^{s^*}/a$ and in the bulk $c_{i,v}^{*}$. In fact, this case corresponds to the infinite ideal crystal in thermodynamic equilibrium. In this case, the surface could be replaced by the intersection of the crystal by the infinite plane. Obviously, the sum of the point defect fluxes through this plane is equal to zero in equilibrium conditions and corresponding energy barriers for the point defect penetration through the plane are equivalent. 

Let us consider the obtained result for silicon --- the most investigated semiconductor.  According to number of studies the formation energies for self-interstitials and bulk vacancies are $E_{i}^{f}=$3.18--3.5\,eV and $E_{v}^{f}=$3.36--3,95\,eV, respectively~\cite{seebauer2006control,kube2013contributions,sudkamp2016self}. The corresponding activation energies for adatom and advacancy formation at Si(111) surface are estimated as $E_{i}^{ad}=0.23$\,eV~\cite{pang2008step}, and $E_{v}^{ad}=1.5\pm0.15$\,eV~\cite{sitnikov2015attachment}. Finally, for slef-interstitials and vacancies we obtain  $E_{i}^{p+ps} - E_{i}^{ads+ps\prime}  = 2.95 \div 3.27~eV$, and $E_{v}^{p+ps} - E_{v}^{ads+ps\prime} = 1,71 \div 2.60~eV$, respectively.
According to Ref.~\cite{sitnikov2017advacancy} the activation energy for vacancy dissolving from (111) silicon surface into the bulk is $4.3\pm0.05$\,eV. Taking this value as $E_{v}^{p+ps}$ we can estimate the energy barrier for vacancy penetration to the surface $E_{v}^{ads+ps\prime}=1.65 \div 2.59$\,eV.  Recently, the energy barrier for the penetration of self-interstitials from bulk to the surface has been estimated as  $E_{i}^{ads+ps\prime}=1.5\pm0.2$\,eV~\cite{kosolobov2019real}. Thus, we can estimate the total energy barrier for the self-interstitials dissolving from the surface to bulk $E_{i}^{p+ps}=4.25 \div 4.97$\,eV.

\begin{figure}
	\includegraphics[width=7.5cm]{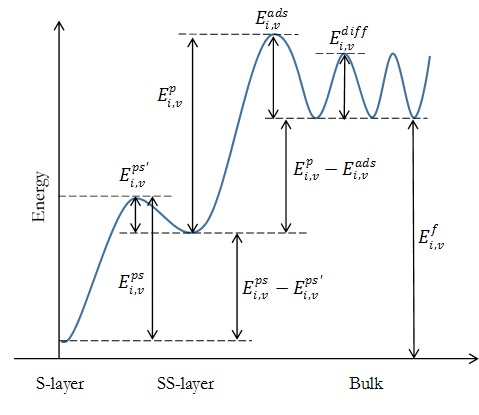}
	\caption{\label{fig-scheme} Schematic representation of potential energy diagram for the diffusion of point defects  from the surface to subsurface layer and in the bulk. }
\end{figure}

Figure~\ref{fig-scheme} shows the schematic representation of the energy barriers in a crystal near the crystal-vacuum boundary. The positions of the surface layer (S-layer), subsurface (SS-layer), and bulk are indicated at the ordinate axis. As it is seen from the picture the energy difference $E_{i,v}^{ps}-E_{i,v}^{ps \prime}$ reflects the position of the energy level for the subsurface layer with respect to the surface layer potential, whereas   $E_{i,v}^{p}-E_{i,v}^{ads}$ shows the difference in potentials between the subsurface layer and bulk level. The process of bulk diffusion is characterized by $E_{i,v}^{dif}$ energy barrier, depicted on the right side of the figure, $E_{i,v}^{f}$  is the bulk formation energy of the corresponding defect.

The potential profile shown in Fig.~\ref{fig-scheme} qualitatively agreed with the density functional theory predictions for the silicon adatom diffusion from the surface to subsurface layers reported in~\cite{uberuaga2000diffusion}. 
The dependence of the point defect formation energy on the depth near silicon and germanium surfaces also reported~\cite{kamiyama2012first}. By numerical simulations, it was shown that formation energies of vacancies and self-interstitials near crystal surface are lower than that of defects in the bulk of crystal. Self-interstitials can form easier than vacancies near the surface boundary and the formation energy of the self-interstitials and vacancies increase when moving inside the crystal. For vacancies in the first atomic layer, the formation energy is about 2.4\,eV lower than $E_{v}^{f}$ in the sixth layer. Whereas for the self-interstitial the difference in $E_{i}^{f}$ for the sixth and second atomic layer is about 2.7\,eV. The reported values well correspond to the predictions of our model. 
Seebauer~\cite{seebauer2006control} showed that defect concentration in the subsurface layers of silicon crystals can be profoundly modified by gas adsorption. The concentration of point defect in silicon samples exposed to low coverages of the nitrogen at temperatures around 800$^{\circ}$C varied over several orders of magnitude.      
Recently, it was shown with the use of \textit{in situ} atomic-resolution electron microscopy that atomic steps affect the composition and structure of the subsurface atomic layers of metal alloys~\cite{liu2021}. These findings support the predictions that the surface processes induce the changes in bulk diffusion at least in the subsurface region. The proposed model includes the formation of the subsurface layer, containing self-interstitials and adsorbed bulk vacancies diffusing along the crystal-vacuum boundary below the surface. The formation of this layer provides the opportunity to maintain the equilibrium concentrations of point defects at the crystal boundary by producing or adsorbing the bulk vacancies and self-interstitial by atomic steps. This mechanism not only explains the ability of the crystal surfaces to produce or consume bulk point defects but also demonstrates the link between surface and bulk diffusion processes.

In summary, the behavior of the native point defects near the crystal-vacuum boundary has been investigated theoretically. The system of self-contained differential equations is proposed to describe the point defect formation and diffusion in crystal bulk and at the surface boundary. The results show the fundamental relation between the equilibrium point defect concentrations in bulk and at the surface of the crystal. Moreover, the formation energy of the bulk and surface point defects are linked through the energy barriers that restricted the fluxes of point defects between the surface and bulk of the crystal. Beyond the fundamental relevance, the simple and general nature of the proposed model suggests a universal approach for the prediction of the point defect behavior in crystals and nanoscale structures in various solid state systems.

\bibliography{PRL-coupled_02}

\begin{thebibliography}{32}%
\makeatletter
\providecommand \@ifxundefined [1]{%
 \@ifx{#1\undefined}
}%
\providecommand \@ifnum [1]{%
 \ifnum #1\expandafter \@firstoftwo
 \else \expandafter \@secondoftwo
 \fi
}%
\providecommand \@ifx [1]{%
 \ifx #1\expandafter \@firstoftwo
 \else \expandafter \@secondoftwo
 \fi
}%
\providecommand \natexlab [1]{#1}%
\providecommand \enquote  [1]{``#1''}%
\providecommand \bibnamefont  [1]{#1}%
\providecommand \bibfnamefont [1]{#1}%
\providecommand \citenamefont [1]{#1}%
\providecommand \href@noop [0]{\@secondoftwo}%
\providecommand \href [0]{\begingroup \@sanitize@url \@href}%
\providecommand \@href[1]{\@@startlink{#1}\@@href}%
\providecommand \@@href[1]{\endgroup#1\@@endlink}%
\providecommand \@sanitize@url [0]{\catcode `\\12\catcode `\$12\catcode
  `\&12\catcode `\#12\catcode `\^12\catcode `\_12\catcode `\%12\relax}%
\providecommand \@@startlink[1]{}%
\providecommand \@@endlink[0]{}%
\providecommand \url  [0]{\begingroup\@sanitize@url \@url }%
\providecommand \@url [1]{\endgroup\@href {#1}{\urlprefix }}%
\providecommand \urlprefix  [0]{URL }%
\providecommand \Eprint [0]{\href }%
\providecommand \doibase [0]{https://doi.org/}%
\providecommand \selectlanguage [0]{\@gobble}%
\providecommand \bibinfo  [0]{\@secondoftwo}%
\providecommand \bibfield  [0]{\@secondoftwo}%
\providecommand \translation [1]{[#1]}%
\providecommand \BibitemOpen [0]{}%
\providecommand \bibitemStop [0]{}%
\providecommand \bibitemNoStop [0]{.\EOS\space}%
\providecommand \EOS [0]{\spacefactor3000\relax}%
\providecommand \BibitemShut  [1]{\csname bibitem#1\endcsname}%
\let\auto@bib@innerbib\@empty
\bibitem [{\citenamefont {Pantelides}(1978)}]{pantelides1978electronic}%
  \BibitemOpen
  \bibfield  {author} {\bibinfo {author} {\bibfnamefont {S.~T.}\ \bibnamefont
  {Pantelides}},\ }\href@noop {} {\bibfield  {journal} {\bibinfo  {journal}
  {Reviews of Modern Physics}\ }\textbf {\bibinfo {volume} {50}},\ \bibinfo
  {pages} {797} (\bibinfo {year} {1978})}\BibitemShut {NoStop}%
\bibitem [{\citenamefont {Yoshida}\ and\ \citenamefont
  {Langouche}(2015)}]{yoshida2015defects}%
  \BibitemOpen
  \bibfield  {author} {\bibinfo {author} {\bibfnamefont {Y.}~\bibnamefont
  {Yoshida}}\ and\ \bibinfo {author} {\bibfnamefont {G.}~\bibnamefont
  {Langouche}},\ }\href@noop {} {\emph {\bibinfo {title} {Defects and
  Impurities in Silicon Materials}}}\ (\bibinfo  {publisher} {Springer},\
  \bibinfo {year} {2015})\BibitemShut {NoStop}%
\bibitem [{\citenamefont {Koenraad}\ and\ \citenamefont
  {Flatt{\'e}}(2011)}]{koenraad2011single}%
  \BibitemOpen
  \bibfield  {author} {\bibinfo {author} {\bibfnamefont {P.~M.}\ \bibnamefont
  {Koenraad}}\ and\ \bibinfo {author} {\bibfnamefont {M.~E.}\ \bibnamefont
  {Flatt{\'e}}},\ }\href@noop {} {\bibfield  {journal} {\bibinfo  {journal}
  {Nature materials}\ }\textbf {\bibinfo {volume} {10}},\ \bibinfo {pages} {91}
  (\bibinfo {year} {2011})}\BibitemShut {NoStop}%
\bibitem [{\citenamefont {Park}\ \emph {et~al.}(2018)\citenamefont {Park},
  \citenamefont {Kim}, \citenamefont {Xie},\ and\ \citenamefont
  {Walsh}}]{park2018point}%
  \BibitemOpen
  \bibfield  {author} {\bibinfo {author} {\bibfnamefont {J.~S.}\ \bibnamefont
  {Park}}, \bibinfo {author} {\bibfnamefont {S.}~\bibnamefont {Kim}}, \bibinfo
  {author} {\bibfnamefont {Z.}~\bibnamefont {Xie}},\ and\ \bibinfo {author}
  {\bibfnamefont {A.}~\bibnamefont {Walsh}},\ }\href@noop {} {\bibfield
  {journal} {\bibinfo  {journal} {Nature Reviews Materials}\ }\textbf {\bibinfo
  {volume} {3}},\ \bibinfo {pages} {194} (\bibinfo {year} {2018})}\BibitemShut
  {NoStop}%
\bibitem [{\citenamefont {Watanabe}\ \emph {et~al.}(2004)\citenamefont
  {Watanabe}, \citenamefont {Taniguchi},\ and\ \citenamefont
  {Kanda}}]{watanabe2004direct}%
  \BibitemOpen
  \bibfield  {author} {\bibinfo {author} {\bibfnamefont {K.}~\bibnamefont
  {Watanabe}}, \bibinfo {author} {\bibfnamefont {T.}~\bibnamefont
  {Taniguchi}},\ and\ \bibinfo {author} {\bibfnamefont {H.}~\bibnamefont
  {Kanda}},\ }\href@noop {} {\bibfield  {journal} {\bibinfo  {journal} {Nat.
  Mater.}\ }\textbf {\bibinfo {volume} {3}},\ \bibinfo {pages} {404} (\bibinfo
  {year} {2004})}\BibitemShut {NoStop}%
\bibitem [{\citenamefont {Boyes}\ \emph {et~al.}(1985)\citenamefont {Boyes},
  \citenamefont {Gai},\ and\ \citenamefont
  {Warwick}}]{boyes1985cathodoluminescence}%
  \BibitemOpen
  \bibfield  {author} {\bibinfo {author} {\bibfnamefont {E.}~\bibnamefont
  {Boyes}}, \bibinfo {author} {\bibfnamefont {P.}~\bibnamefont {Gai}},\ and\
  \bibinfo {author} {\bibfnamefont {C.}~\bibnamefont {Warwick}},\ }\href@noop
  {} {\bibfield  {journal} {\bibinfo  {journal} {Nature}\ }\textbf {\bibinfo
  {volume} {313}},\ \bibinfo {pages} {666} (\bibinfo {year}
  {1985})}\BibitemShut {NoStop}%
\bibitem [{\citenamefont {Ledentu}\ \emph {et~al.}(2000)\citenamefont
  {Ledentu}, \citenamefont {Dong},\ and\ \citenamefont
  {Sautet}}]{ledentu2000heterogeneous}%
  \BibitemOpen
  \bibfield  {author} {\bibinfo {author} {\bibfnamefont {V.}~\bibnamefont
  {Ledentu}}, \bibinfo {author} {\bibfnamefont {W.}~\bibnamefont {Dong}},\ and\
  \bibinfo {author} {\bibfnamefont {P.}~\bibnamefont {Sautet}},\ }\href@noop {}
  {\bibfield  {journal} {\bibinfo  {journal} {J. Am. Chem. Soc.}\ }\textbf
  {\bibinfo {volume} {122}},\ \bibinfo {pages} {1796} (\bibinfo {year}
  {2000})}\BibitemShut {NoStop}%
\bibitem [{\citenamefont {Toijer}\ \emph {et~al.}(2021)\citenamefont {Toijer},
  \citenamefont {Messina}, \citenamefont {Domain}, \citenamefont {Vidal},
  \citenamefont {Becquart},\ and\ \citenamefont {Olsson}}]{toijer2021solute}%
  \BibitemOpen
  \bibfield  {author} {\bibinfo {author} {\bibfnamefont {E.}~\bibnamefont
  {Toijer}}, \bibinfo {author} {\bibfnamefont {L.}~\bibnamefont {Messina}},
  \bibinfo {author} {\bibfnamefont {C.}~\bibnamefont {Domain}}, \bibinfo
  {author} {\bibfnamefont {J.}~\bibnamefont {Vidal}}, \bibinfo {author}
  {\bibfnamefont {C.}~\bibnamefont {Becquart}},\ and\ \bibinfo {author}
  {\bibfnamefont {P.}~\bibnamefont {Olsson}},\ }\href@noop {} {\bibfield
  {journal} {\bibinfo  {journal} {Phys. Rev. Mater.}\ }\textbf {\bibinfo
  {volume} {5}},\ \bibinfo {pages} {013602} (\bibinfo {year}
  {2021})}\BibitemShut {NoStop}%
\bibitem [{\citenamefont {Zhang}\ \emph {et~al.}(2018)\citenamefont {Zhang},
  \citenamefont {Hattar}, \citenamefont {Chen}, \citenamefont {Shao},
  \citenamefont {Li}, \citenamefont {Sun}, \citenamefont {Yu}, \citenamefont
  {Li}, \citenamefont {Taheri}, \citenamefont {Wang} \emph
  {et~al.}}]{zhang2018radiation}%
  \BibitemOpen
  \bibfield  {author} {\bibinfo {author} {\bibfnamefont {X.}~\bibnamefont
  {Zhang}}, \bibinfo {author} {\bibfnamefont {K.}~\bibnamefont {Hattar}},
  \bibinfo {author} {\bibfnamefont {Y.}~\bibnamefont {Chen}}, \bibinfo {author}
  {\bibfnamefont {L.}~\bibnamefont {Shao}}, \bibinfo {author} {\bibfnamefont
  {J.}~\bibnamefont {Li}}, \bibinfo {author} {\bibfnamefont {C.}~\bibnamefont
  {Sun}}, \bibinfo {author} {\bibfnamefont {K.}~\bibnamefont {Yu}}, \bibinfo
  {author} {\bibfnamefont {N.}~\bibnamefont {Li}}, \bibinfo {author}
  {\bibfnamefont {M.~L.}\ \bibnamefont {Taheri}}, \bibinfo {author}
  {\bibfnamefont {H.}~\bibnamefont {Wang}}, \emph {et~al.},\ }\href@noop {}
  {\bibfield  {journal} {\bibinfo  {journal} {Prog. Mater. Sci.}\ }\textbf
  {\bibinfo {volume} {96}},\ \bibinfo {pages} {217} (\bibinfo {year}
  {2018})}\BibitemShut {NoStop}%
\bibitem [{\citenamefont {Holmberg}\ \emph {et~al.}(2013)\citenamefont
  {Holmberg}, \citenamefont {Helps}, \citenamefont {Mkhoyan},\ and\
  \citenamefont {Norris}}]{holmberg2013imaging}%
  \BibitemOpen
  \bibfield  {author} {\bibinfo {author} {\bibfnamefont {V.~C.}\ \bibnamefont
  {Holmberg}}, \bibinfo {author} {\bibfnamefont {J.~R.}\ \bibnamefont {Helps}},
  \bibinfo {author} {\bibfnamefont {K.~A.}\ \bibnamefont {Mkhoyan}},\ and\
  \bibinfo {author} {\bibfnamefont {D.~J.}\ \bibnamefont {Norris}},\
  }\href@noop {} {\bibfield  {journal} {\bibinfo  {journal} {Chem. Mater.}\
  }\textbf {\bibinfo {volume} {25}},\ \bibinfo {pages} {1332} (\bibinfo {year}
  {2013})}\BibitemShut {NoStop}%
\bibitem [{\citenamefont {Sakurai}\ \emph {et~al.}(2012)\citenamefont
  {Sakurai}, \citenamefont {Nishino}, \citenamefont {Futaba}, \citenamefont
  {Yasuda}, \citenamefont {Yamada}, \citenamefont {Maigne}, \citenamefont
  {Matsuo}, \citenamefont {Nakamura}, \citenamefont {Yumura},\ and\
  \citenamefont {Hata}}]{sakurai2012role}%
  \BibitemOpen
  \bibfield  {author} {\bibinfo {author} {\bibfnamefont {S.}~\bibnamefont
  {Sakurai}}, \bibinfo {author} {\bibfnamefont {H.}~\bibnamefont {Nishino}},
  \bibinfo {author} {\bibfnamefont {D.~N.}\ \bibnamefont {Futaba}}, \bibinfo
  {author} {\bibfnamefont {S.}~\bibnamefont {Yasuda}}, \bibinfo {author}
  {\bibfnamefont {T.}~\bibnamefont {Yamada}}, \bibinfo {author} {\bibfnamefont
  {A.}~\bibnamefont {Maigne}}, \bibinfo {author} {\bibfnamefont
  {Y.}~\bibnamefont {Matsuo}}, \bibinfo {author} {\bibfnamefont
  {E.}~\bibnamefont {Nakamura}}, \bibinfo {author} {\bibfnamefont
  {M.}~\bibnamefont {Yumura}},\ and\ \bibinfo {author} {\bibfnamefont
  {K.}~\bibnamefont {Hata}},\ }\href@noop {} {\bibfield  {journal} {\bibinfo
  {journal} {J. Am. Chem. Soc.}\ }\textbf {\bibinfo {volume} {134}},\ \bibinfo
  {pages} {2148} (\bibinfo {year} {2012})}\BibitemShut {NoStop}%
\bibitem [{\citenamefont {Beyerlein}\ \emph {et~al.}(2015)\citenamefont
  {Beyerlein}, \citenamefont {Demkowicz}, \citenamefont {Misra},\ and\
  \citenamefont {Uberuaga}}]{beyerlein2015defect}%
  \BibitemOpen
  \bibfield  {author} {\bibinfo {author} {\bibfnamefont {I.~J.}\ \bibnamefont
  {Beyerlein}}, \bibinfo {author} {\bibfnamefont {M.~J.}\ \bibnamefont
  {Demkowicz}}, \bibinfo {author} {\bibfnamefont {A.}~\bibnamefont {Misra}},\
  and\ \bibinfo {author} {\bibfnamefont {B.}~\bibnamefont {Uberuaga}},\
  }\href@noop {} {\bibfield  {journal} {\bibinfo  {journal} {Prog. Mater.
  Sci.}\ }\textbf {\bibinfo {volume} {74}},\ \bibinfo {pages} {125} (\bibinfo
  {year} {2015})}\BibitemShut {NoStop}%
\bibitem [{\citenamefont {Pichler}(2004)}]{pichler_book2004}%
  \BibitemOpen
  \bibfield  {author} {\bibinfo {author} {\bibfnamefont {P.}~\bibnamefont
  {Pichler}},\ }\bibinfo {title} {Intrinsic point defects, impurities, and
  their diffusion in silicon}\ (\bibinfo  {publisher} {Springer, Vienna.},\
  \bibinfo {year} {2004})\ Chap.\ \bibinfo {chapter} {Intrinsic Point Defects},
  pp.\ \bibinfo {pages} {77--227}\BibitemShut {NoStop}%
\bibitem [{\citenamefont {Kosolobov}(2019)}]{kosolobov2019subsurface}%
  \BibitemOpen
  \bibfield  {author} {\bibinfo {author} {\bibfnamefont {S.}~\bibnamefont
  {Kosolobov}},\ }\href@noop {} {\bibfield  {journal} {\bibinfo  {journal}
  {Scientific Reports}\ }\textbf {\bibinfo {volume} {9}},\ \bibinfo {pages} {1}
  (\bibinfo {year} {2019})}\BibitemShut {NoStop}%
\bibitem [{\citenamefont {Fahey}\ \emph {et~al.}(1989)\citenamefont {Fahey},
  \citenamefont {Griffin},\ and\ \citenamefont {Plummer}}]{fahey1989}%
  \BibitemOpen
  \bibfield  {author} {\bibinfo {author} {\bibfnamefont {P.~M.}\ \bibnamefont
  {Fahey}}, \bibinfo {author} {\bibfnamefont {P.~B.}\ \bibnamefont {Griffin}},\
  and\ \bibinfo {author} {\bibfnamefont {J.~D.}\ \bibnamefont {Plummer}},\
  }\href {https://doi.org/10.1103/RevModPhys.61.289} {\bibfield  {journal}
  {\bibinfo  {journal} {Rev. Mod. Phys.}\ }\textbf {\bibinfo {volume} {61}},\
  \bibinfo {pages} {289} (\bibinfo {year} {1989})}\BibitemShut {NoStop}%
\bibitem [{\citenamefont {Hu}(1994)}]{hu1994nonequilibrium}%
  \BibitemOpen
  \bibfield  {author} {\bibinfo {author} {\bibfnamefont {S.}~\bibnamefont
  {Hu}},\ }\href@noop {} {\bibfield  {journal} {\bibinfo  {journal} {Mater.
  Sci. Eng.}\ }\textbf {\bibinfo {volume} {13}},\ \bibinfo {pages} {105}
  (\bibinfo {year} {1994})}\BibitemShut {NoStop}%
\bibitem [{\citenamefont {Mikelsen}\ \emph {et~al.}(2005)\citenamefont
  {Mikelsen}, \citenamefont {Monakhov}, \citenamefont {Alfieri}, \citenamefont
  {Avset},\ and\ \citenamefont {Svensson}}]{mikelsen2005}%
  \BibitemOpen
  \bibfield  {author} {\bibinfo {author} {\bibfnamefont {M.}~\bibnamefont
  {Mikelsen}}, \bibinfo {author} {\bibfnamefont {E.~V.}\ \bibnamefont
  {Monakhov}}, \bibinfo {author} {\bibfnamefont {G.}~\bibnamefont {Alfieri}},
  \bibinfo {author} {\bibfnamefont {B.~S.}\ \bibnamefont {Avset}},\ and\
  \bibinfo {author} {\bibfnamefont {B.~G.}\ \bibnamefont {Svensson}},\
  }\href@noop {} {\bibfield  {journal} {\bibinfo  {journal} {Phys. Rev. B}\
  }\textbf {\bibinfo {volume} {72}},\ \bibinfo {pages} {195207} (\bibinfo
  {year} {2005})}\BibitemShut {NoStop}%
\bibitem [{\citenamefont {Watkins}(2008)}]{watkins2008vacancy}%
  \BibitemOpen
  \bibfield  {author} {\bibinfo {author} {\bibfnamefont {G.~D.}\ \bibnamefont
  {Watkins}},\ }\href@noop {} {\bibfield  {journal} {\bibinfo  {journal} {J.
  Appl. Phys.}\ }\textbf {\bibinfo {volume} {103}},\ \bibinfo {pages} {106106}
  (\bibinfo {year} {2008})}\BibitemShut {NoStop}%
\bibitem [{\citenamefont {Bracht}\ \emph {et~al.}(2003)\citenamefont {Bracht},
  \citenamefont {Pedersen}, \citenamefont {Zangenberg}, \citenamefont {Larsen},
  \citenamefont {Haller}, \citenamefont {Lulli},\ and\ \citenamefont
  {Posselt}}]{bracht2003radiation}%
  \BibitemOpen
  \bibfield  {author} {\bibinfo {author} {\bibfnamefont {H.}~\bibnamefont
  {Bracht}}, \bibinfo {author} {\bibfnamefont {J.~F.}\ \bibnamefont
  {Pedersen}}, \bibinfo {author} {\bibfnamefont {N.}~\bibnamefont
  {Zangenberg}}, \bibinfo {author} {\bibfnamefont {A.~N.}\ \bibnamefont
  {Larsen}}, \bibinfo {author} {\bibfnamefont {E.~E.}\ \bibnamefont {Haller}},
  \bibinfo {author} {\bibfnamefont {G.}~\bibnamefont {Lulli}},\ and\ \bibinfo
  {author} {\bibfnamefont {M.}~\bibnamefont {Posselt}},\ }\href@noop {}
  {\bibfield  {journal} {\bibinfo  {journal} {Phys. Rev. Lett.}\ }\textbf
  {\bibinfo {volume} {91}},\ \bibinfo {pages} {245502} (\bibinfo {year}
  {2003})}\BibitemShut {NoStop}%
\bibitem [{\citenamefont {Kissinger}\ \emph {et~al.}(2014)\citenamefont
  {Kissinger}, \citenamefont {Dabrowski},\ and\ \citenamefont
  {Kot}}]{kissinger2014simulation}%
  \BibitemOpen
  \bibfield  {author} {\bibinfo {author} {\bibfnamefont {G.}~\bibnamefont
  {Kissinger}}, \bibinfo {author} {\bibfnamefont {J.}~\bibnamefont
  {Dabrowski}},\ and\ \bibinfo {author} {\bibfnamefont {D.}~\bibnamefont
  {Kot}},\ }\href@noop {} {\bibfield  {journal} {\bibinfo  {journal} {Jap. J.
  Appl. Phys.}\ }\textbf {\bibinfo {volume} {53}},\ \bibinfo {pages} {05FJ06}
  (\bibinfo {year} {2014})}\BibitemShut {NoStop}%
\bibitem [{\citenamefont {Man}\ \emph {et~al.}(2007)\citenamefont {Man},
  \citenamefont {Pang},\ and\ \citenamefont {Altman}}]{man2007kinetic}%
  \BibitemOpen
  \bibfield  {author} {\bibinfo {author} {\bibfnamefont {K.}~\bibnamefont
  {Man}}, \bibinfo {author} {\bibfnamefont {A.}~\bibnamefont {Pang}},\ and\
  \bibinfo {author} {\bibfnamefont {M.}~\bibnamefont {Altman}},\ }\href@noop {}
  {\bibfield  {journal} {\bibinfo  {journal} {Surf. Sci.}\ }\textbf {\bibinfo
  {volume} {601}},\ \bibinfo {pages} {4669} (\bibinfo {year}
  {2007})}\BibitemShut {NoStop}%
\bibitem [{\citenamefont {Bales}\ and\ \citenamefont
  {Zangwill}(1997)}]{bales1997}%
  \BibitemOpen
  \bibfield  {author} {\bibinfo {author} {\bibfnamefont {G.~S.}\ \bibnamefont
  {Bales}}\ and\ \bibinfo {author} {\bibfnamefont {A.}~\bibnamefont
  {Zangwill}},\ }\href@noop {} {\bibfield  {journal} {\bibinfo  {journal}
  {Phys. Rev. B}\ }\textbf {\bibinfo {volume} {55}},\ \bibinfo {pages} {R1973}
  (\bibinfo {year} {1997})}\BibitemShut {NoStop}%
\bibitem [{\citenamefont {Seebauer}\ \emph {et~al.}(2006)\citenamefont
  {Seebauer}, \citenamefont {Dev}, \citenamefont {Jung}, \citenamefont
  {Vaidyanathan}, \citenamefont {Kwok}, \citenamefont {Ager}, \citenamefont
  {Haller},\ and\ \citenamefont {Braatz}}]{seebauer2006control}%
  \BibitemOpen
  \bibfield  {author} {\bibinfo {author} {\bibfnamefont {E.~G.}\ \bibnamefont
  {Seebauer}}, \bibinfo {author} {\bibfnamefont {K.}~\bibnamefont {Dev}},
  \bibinfo {author} {\bibfnamefont {M.~Y.}\ \bibnamefont {Jung}}, \bibinfo
  {author} {\bibfnamefont {R.}~\bibnamefont {Vaidyanathan}}, \bibinfo {author}
  {\bibfnamefont {C.~T.}\ \bibnamefont {Kwok}}, \bibinfo {author}
  {\bibfnamefont {J.~W.}\ \bibnamefont {Ager}}, \bibinfo {author}
  {\bibfnamefont {E.~E.}\ \bibnamefont {Haller}},\ and\ \bibinfo {author}
  {\bibfnamefont {R.~D.}\ \bibnamefont {Braatz}},\ }\href@noop {} {\bibfield
  {journal} {\bibinfo  {journal} {Phys. Rev. Lett.}\ }\textbf {\bibinfo
  {volume} {97}},\ \bibinfo {pages} {055503} (\bibinfo {year}
  {2006})}\BibitemShut {NoStop}%
\bibitem [{\citenamefont {Kube}\ \emph {et~al.}(2013)\citenamefont {Kube},
  \citenamefont {Bracht}, \citenamefont {H\"uger}, \citenamefont {Schmidt},
  \citenamefont {Hansen}, \citenamefont {Larsen}, \citenamefont {Ager},
  \citenamefont {Haller}, \citenamefont {Geue},\ and\ \citenamefont
  {Stahn}}]{kube2013contributions}%
  \BibitemOpen
  \bibfield  {author} {\bibinfo {author} {\bibfnamefont {R.}~\bibnamefont
  {Kube}}, \bibinfo {author} {\bibfnamefont {H.}~\bibnamefont {Bracht}},
  \bibinfo {author} {\bibfnamefont {E.}~\bibnamefont {H\"uger}}, \bibinfo
  {author} {\bibfnamefont {H.}~\bibnamefont {Schmidt}}, \bibinfo {author}
  {\bibfnamefont {J.~L.}\ \bibnamefont {Hansen}}, \bibinfo {author}
  {\bibfnamefont {A.~N.}\ \bibnamefont {Larsen}}, \bibinfo {author}
  {\bibfnamefont {J.~W.}\ \bibnamefont {Ager}}, \bibinfo {author}
  {\bibfnamefont {E.~E.}\ \bibnamefont {Haller}}, \bibinfo {author}
  {\bibfnamefont {T.}~\bibnamefont {Geue}},\ and\ \bibinfo {author}
  {\bibfnamefont {J.}~\bibnamefont {Stahn}},\ }\href@noop {} {\bibfield
  {journal} {\bibinfo  {journal} {Phys. Rev. B}\ }\textbf {\bibinfo {volume}
  {88}},\ \bibinfo {pages} {085206} (\bibinfo {year} {2013})}\BibitemShut
  {NoStop}%
\bibitem [{\citenamefont {S\"udkamp}\ and\ \citenamefont
  {Bracht}(2016)}]{sudkamp2016self}%
  \BibitemOpen
  \bibfield  {author} {\bibinfo {author} {\bibfnamefont {T.}~\bibnamefont
  {S\"udkamp}}\ and\ \bibinfo {author} {\bibfnamefont {H.}~\bibnamefont
  {Bracht}},\ }\href@noop {} {\bibfield  {journal} {\bibinfo  {journal} {Phys.
  Rev. B}\ }\textbf {\bibinfo {volume} {94}},\ \bibinfo {pages} {125208}
  (\bibinfo {year} {2016})}\BibitemShut {NoStop}%
\bibitem [{\citenamefont {Pang}\ \emph {et~al.}(2008)\citenamefont {Pang},
  \citenamefont {Man}, \citenamefont {Altman}, \citenamefont {Stasevich},
  \citenamefont {Szalma},\ and\ \citenamefont {Einstein}}]{pang2008step}%
  \BibitemOpen
  \bibfield  {author} {\bibinfo {author} {\bibfnamefont {A.}~\bibnamefont
  {Pang}}, \bibinfo {author} {\bibfnamefont {K.}~\bibnamefont {Man}}, \bibinfo
  {author} {\bibfnamefont {M.~S.}\ \bibnamefont {Altman}}, \bibinfo {author}
  {\bibfnamefont {T.}~\bibnamefont {Stasevich}}, \bibinfo {author}
  {\bibfnamefont {F.}~\bibnamefont {Szalma}},\ and\ \bibinfo {author}
  {\bibfnamefont {T.}~\bibnamefont {Einstein}},\ }\href@noop {} {\bibfield
  {journal} {\bibinfo  {journal} {Phys. Rev. B}\ }\textbf {\bibinfo {volume}
  {77}},\ \bibinfo {pages} {115424} (\bibinfo {year} {2008})}\BibitemShut
  {NoStop}%
\bibitem [{\citenamefont {Sitnikov}\ \emph {et~al.}(2015)\citenamefont
  {Sitnikov}, \citenamefont {Kosolobov},\ and\ \citenamefont
  {Latyshev}}]{sitnikov2015attachment}%
  \BibitemOpen
  \bibfield  {author} {\bibinfo {author} {\bibfnamefont {S.}~\bibnamefont
  {Sitnikov}}, \bibinfo {author} {\bibfnamefont {S.}~\bibnamefont
  {Kosolobov}},\ and\ \bibinfo {author} {\bibfnamefont {A.}~\bibnamefont
  {Latyshev}},\ }\href@noop {} {\bibfield  {journal} {\bibinfo  {journal}
  {Surf. Sci.}\ }\textbf {\bibinfo {volume} {633}},\ \bibinfo {pages} {L1}
  (\bibinfo {year} {2015})}\BibitemShut {NoStop}%
\bibitem [{\citenamefont {Sitnikov}\ \emph {et~al.}(2017)\citenamefont
  {Sitnikov}, \citenamefont {Latyshev},\ and\ \citenamefont
  {Kosolobov}}]{sitnikov2017advacancy}%
  \BibitemOpen
  \bibfield  {author} {\bibinfo {author} {\bibfnamefont {S.}~\bibnamefont
  {Sitnikov}}, \bibinfo {author} {\bibfnamefont {A.}~\bibnamefont {Latyshev}},\
  and\ \bibinfo {author} {\bibfnamefont {S.}~\bibnamefont {Kosolobov}},\
  }\href@noop {} {\bibfield  {journal} {\bibinfo  {journal} {J. Cryst. Growth}\
  }\textbf {\bibinfo {volume} {457}},\ \bibinfo {pages} {196} (\bibinfo {year}
  {2017})}\BibitemShut {NoStop}%
\bibitem [{\citenamefont {Kosolobov}\ \emph {et~al.}(2019)\citenamefont
  {Kosolobov}, \citenamefont {Nazarikov}, \citenamefont {Sitnikov},
  \citenamefont {Pshenichnyuk}, \citenamefont {Fedina},\ and\ \citenamefont
  {Latyshev}}]{kosolobov2019real}%
  \BibitemOpen
  \bibfield  {author} {\bibinfo {author} {\bibfnamefont {S.}~\bibnamefont
  {Kosolobov}}, \bibinfo {author} {\bibfnamefont {G.}~\bibnamefont
  {Nazarikov}}, \bibinfo {author} {\bibfnamefont {S.}~\bibnamefont {Sitnikov}},
  \bibinfo {author} {\bibfnamefont {I.}~\bibnamefont {Pshenichnyuk}}, \bibinfo
  {author} {\bibfnamefont {L.}~\bibnamefont {Fedina}},\ and\ \bibinfo {author}
  {\bibfnamefont {A.}~\bibnamefont {Latyshev}},\ }\href@noop {} {\bibfield
  {journal} {\bibinfo  {journal} {Surf. Sci.}\ }\textbf {\bibinfo {volume}
  {687}},\ \bibinfo {pages} {25} (\bibinfo {year} {2019})}\BibitemShut
  {NoStop}%
\bibitem [{\citenamefont {Uberuaga}\ \emph {et~al.}(2000)\citenamefont
  {Uberuaga}, \citenamefont {Leskovar}, \citenamefont {Smith}, \citenamefont
  {J{\'o}nsson},\ and\ \citenamefont {Olmstead}}]{uberuaga2000diffusion}%
  \BibitemOpen
  \bibfield  {author} {\bibinfo {author} {\bibfnamefont {B.~P.}\ \bibnamefont
  {Uberuaga}}, \bibinfo {author} {\bibfnamefont {M.}~\bibnamefont {Leskovar}},
  \bibinfo {author} {\bibfnamefont {A.~P.}\ \bibnamefont {Smith}}, \bibinfo
  {author} {\bibfnamefont {H.}~\bibnamefont {J{\'o}nsson}},\ and\ \bibinfo
  {author} {\bibfnamefont {M.}~\bibnamefont {Olmstead}},\ }\href@noop {}
  {\bibfield  {journal} {\bibinfo  {journal} {Phys. Rev. Lett.}\ }\textbf
  {\bibinfo {volume} {84}},\ \bibinfo {pages} {2441} (\bibinfo {year}
  {2000})}\BibitemShut {NoStop}%
\bibitem [{\citenamefont {Kamiyama}\ and\ \citenamefont
  {Sueoka}(2012)}]{kamiyama2012first}%
  \BibitemOpen
  \bibfield  {author} {\bibinfo {author} {\bibfnamefont {E.}~\bibnamefont
  {Kamiyama}}\ and\ \bibinfo {author} {\bibfnamefont {K.}~\bibnamefont
  {Sueoka}},\ }\href@noop {} {\bibfield  {journal} {\bibinfo  {journal} {J.
  Appl. Phys.}\ }\textbf {\bibinfo {volume} {111}},\ \bibinfo {pages} {013521}
  (\bibinfo {year} {2012})}\BibitemShut {NoStop}%
\bibitem [{\citenamefont {Liu}\ \emph {et~al.}(2021)\citenamefont {Liu},
  \citenamefont {Zhang}, \citenamefont {Wu}, \citenamefont {Luo}, \citenamefont
  {Sun}, \citenamefont {Chen}, \citenamefont {Zakharov}, \citenamefont {Cheng},
  \citenamefont {Zhu}, \citenamefont {Yang}, \citenamefont {Wang},\ and\
  \citenamefont {Zhou}}]{liu2021}%
  \BibitemOpen
  \bibfield  {author} {\bibinfo {author} {\bibfnamefont {K.}~\bibnamefont
  {Liu}}, \bibinfo {author} {\bibfnamefont {S.}~\bibnamefont {Zhang}}, \bibinfo
  {author} {\bibfnamefont {D.}~\bibnamefont {Wu}}, \bibinfo {author}
  {\bibfnamefont {L.}~\bibnamefont {Luo}}, \bibinfo {author} {\bibfnamefont
  {X.}~\bibnamefont {Sun}}, \bibinfo {author} {\bibfnamefont {X.}~\bibnamefont
  {Chen}}, \bibinfo {author} {\bibfnamefont {D.}~\bibnamefont {Zakharov}},
  \bibinfo {author} {\bibfnamefont {S.}~\bibnamefont {Cheng}}, \bibinfo
  {author} {\bibfnamefont {Y.}~\bibnamefont {Zhu}}, \bibinfo {author}
  {\bibfnamefont {J.~C.}\ \bibnamefont {Yang}}, \bibinfo {author}
  {\bibfnamefont {G.}~\bibnamefont {Wang}},\ and\ \bibinfo {author}
  {\bibfnamefont {G.}~\bibnamefont {Zhou}},\ }\href@noop {} {\bibfield
  {journal} {\bibinfo  {journal} {Phys. Rev. B}\ }\textbf {\bibinfo {volume}
  {103}},\ \bibinfo {pages} {035401} (\bibinfo {year} {2021})}\BibitemShut
  {NoStop}%
\end{thebibliography}%

\end{document}